\documentclass[letterpaper]{jpconf}
\usepackage{graphicx}
\newcommand{\be}{\begin{eqnarray}}
\newcommand{\ee}{\end{eqnarray}}

\newcommand{\Anti}{\mathbf{A}}
\newcommand{\Sym}{\mathbf{S}}

\newcommand{\sixF}{\overline{\mathbf{6}}_F}
\newcommand{\triF}{\mathbf{3}_F}
\newcommand{\sixC}{\overline{\mathbf{6}}_C}
\newcommand{\triC}{\mathbf{3}_C}

\newcommand{\fm}{\mathrm{fm}}
\newcommand{\MeV}{\mathrm{MeV}}

\newcommand{\nf}{N_f}

\def\subs{\underline{S}}
\def\subu{\underline{U}}
\def\subd{\underline{D}}

\begin{document}
\title{
Toward dynamical understanding of the diquarks, pentaquarks and
dibaryons
}

\author{Edward Shuryak}

\address{Department of Physics and Astronomy,\\ State
University of New York at Stony Brook, New York 11794, USA}
\ead{shuryak@tonic.physics.sunysb.edu}

\begin{abstract}
QCD instantons are known to produce deeply bound diquarks. We study
 whether it
 may be used as building blocks in the
formation of multiquark states, in particular pentaquarks and dibaryons.
 A simple model is presented in which the
lowest scalar diquark (and possibly the tensor one)
can be treated as an independent ``body'', with 
the same color  and (approximately) the same mass as a
constituent (anti)quark. In it a new symmetry exists between
states with the same number of ``bodies'' but different number of
quarks appear, in particular the 3-``body'' pentaquarks can be 
naturally related to some excited baryons. 
The limitations of this model are seen from the fact that it leads to
light dibaryon $H$. Another reported work is based on a
 calculation  of a large set of correlation functions for nonlocal operators
with 4 to 6 light quarks in the Random Instanton Liquid Model (RILM).
 The effective interaction between diquarks is 
found to posses a strongly {\em repulsive core}, due to 
the Pauli principle for quark zero modes.

%
\end{abstract}.
\section{Introduction: instantons and diquarks}
The issue of $\bar q q$ and $qq$ basic interaction is at the core of
hadronic
physics. We know that it by no means is restricted to simple one-gluon
exchange
plus confining linear potential,  so familiar from heavy
quarkonia. For light quarks the
instanton-induced interaction \cite{tHooft} plays a very
important or even dominant
role, see \cite{RILM:SS98} for review.
Just to set a scale, recall that the splitting between a pion and
$\eta'$ coming from it: it is of about 800 MeV, the largest 
splitting in hadronic spectroscopy.
 An issue most important for this talk (also discussed in  \cite{RILM:SS98}) is
existence of deeply bound scalar diquarks,  used
for building the color superconducting phases at high density 
\cite{DIQ:BoseCond}.

This talk is however about 
 ``few-body'' problems, in which
 the number of light quarks (+antiquarks) involved is 4,5, and 6. The issue is
especially puzzling at the moment, in view of  sharp contrast between
the
apparent absence of a light dibaryons (see e.g. \cite{H:Review} and
also lattice works such as \cite{H:DESYLat} and others), in sharp
contrast to surprisingly light pentaquark candidates\footnote{
 References/discussion of 
 current experimental status of this issue  is well covered by
many specialized talks at this meeting.}. 
Let me on the onset  underline two general dilemmas one has to face
while discussing the best strategy for building such hadronic states:
(i)   whether 
to follow large $N_c$ (number of colors) ideology
and to focus on  pseudoscalar mesons as clusters,
or the ``small $N_c$ ideology'' and  focus on diquarks.
(ii) the same in a different words:  to use the shell-model
ideology,  so successful for atoms and nuclei, or
to start with the ``pairing'' first;

Let me also emphasize the main point of the talk:
there is a qualitative difference
between perturbative and instanton-based forces.
 One-gluon exchange generates the same interaction
 between two quarks, whatever other quarks do.
Quite  differently, an instanton can serve one quark (per flavor) at
at time only, due to
 Pauli principle for
 't Hooft zero modes.
Thus instantons significantly contribute to clustering
 (both $qq$ and $\bar q q$) at small quark densities only, but are 
much less able to
do so for high density environment.
As will be shown below, this creates quite significant
repulsive interactions between diquarks, reminiscent of the
nuclear core, and rather heavy multiquark states.


The general theoretical reason
 for the lightness of the scalar-isoscalar diquark
state (see e.g.\cite{DIQ:BoseCond}) was known before, it
follows from  Pauli-Gursey 
symmetry of the 2-color QCD. In this theory (the ``small 
$N_c$ limit'' of QCD) the scalar diquarks are actually 
{\it massless Goldstone bosons}. For general $N_c$, the 
instanton (gluon-exchange) in $qq$ is $1/(N_c-1)$ down relative to
$\bar{q}q$. So the real world with $N_c=3$ is half-way between
$N_c=2$ with a relative weight of 1, and $N_c=\infty$ with 
relative weight 0. Loosely speaking, the scalar-isoscalar
diquarks are {\it half Goldstone bosons} with a binding energy of about
$half$ of that for pion, or about one constituent quark mass.

The binding estimate came from a study of
3-quark correlators 
 a decade ago, in instanton liquid \cite{RILM:BaryCor} and
on the lattice~\cite{RILM:BaryLat}. A marked difference between
the nucleon (octet) and $\Delta$ (decuplet) correlators at small times has been
observed, with the former about
 a product of that for quark and a   
 very deeply bound {\em scalar-isoscalar diquark}.
The pseudoscalar channel with $\Gamma=1$ was found to be very strongly
repulsive, the vector and axial vector channels are weakly
repulsive, with a mass of the order of 950 MeV,
above twice the constituent quark mass of the model, 
$2\Sigma=840\, {\rm MeV}$. The only two channels with
attraction and  significant binding are: 
{\bf i.\,} the {\it  scalar} with $m_S\approx\Sigma$ and 
$\Gamma =\gamma_5$; {\bf ii.\,} the {\it  tensor} with
$m_T\approx 570\, {\rm MeV}$ and  $\Gamma=\sigma_{\mu\nu}$ (denoted below
by a subscript $T$).The scalar is odd under spin exchange while the tensor is even under
spin exchange. Fermi statistics forces their flavor to be different.
The scalar is flavor antisymmetric $\bar 3$ while  the tensor
is  flavor symmetric $6$.

\section{A schematic model for pentaquarks based on diquarks}

Because of similar mass and quantum numbers, the
 diquarks may be considered  on equal footing with constituent quarks.
Certain approximate symmetries then appear \cite{PENTA:ShuZah}, relating states
with different number of quarks but 
 the same numbers of ``bodies''. This simple idea 
 is depicted pictorially in Fig.1(left).
The $\bar q q $ mesons (a) are
a well known example of the 2-body objects,
as well as the quark-diquark states (b) (the octet diquark-quark baryons ).
The diquark-antidiquark states (c) are in this model
the 2-body objects.  In $zeroth$ order,
the usual non-strange mesons (like $\rho,\omega$),
   the octet baryons (like the nucleon), and the 4-quark  mesons
(like $a_0(980)$)
are degenerate, with a mass $M\approx2\Sigma= 840\, MeV$.
To  $first$ order, which includes color-related interactions, the  
one-gluon-exchange Coulomb and confinement, the degeneracy should still hold,
as the color charges and the masses of quarks and diquarks are the same.
 Only in $second$ order, when the spin-spin and other residual
 forces are included, they split. 
Note that this new  symmetry 
between $N$, $\rho$ and $a_0(980)$ is more accurate than
the old SU(6) symmetry, which (in zeroth order)
predicts  $M_N\approx M_\Delta$.  

  Pentaquarks and dibaryons  are in this model treated as
3-body objects, with two correlated  diquarks plus 
an antiquark, are thus  
 related to decuplet baryons, see
  Fig.1
(d-f).
 For pentaquarks made of two
scalar diquarks the flavor representations are
$\bar 3\otimes \bar 3 \otimes \bar 3=1\oplus 8\oplus 8\oplus \bar 10$.
Using the   notations with underline for diquarks as contrast to
 bar for antiquarks
where needed,  one can readily see how the pentaquarks observed fit onto
an antidecuplet, $\Theta^+(1540)=(ud)(ud)\bar{s}=\subs \subs \bar{s}$
is an analogue of anti-$\Omega$, and is thus
the top of the antidecuplet (the conjugate of the decuplet). 
New exotic  $\Xi(1860)$ are $\subu \subu \bar u$ and    $\subd \subd
\bar d$, providing the two remaining corners of the triangle. They are
the analogue of anti-$\Delta$.  The remaining 7
members can mix with one octet, as discussed by Jaffe and Wilczek
\cite{PENTA:JafWil}.
making together 18 states in  flavor representations
$({\bf 8}\oplus {\bf {\overline 10}})$.
For ordinary 3 quarks there is the overall Fermi statistics
which ties together flavor and spin-space symmetry and works against
the remaining $1\oplus 8$. There is no such argument for
pentaquarks. So how are the additional flavor states 
$1\oplus 8$ excluded for pentaquarks?
 
For diquark-diquark-antiquark all there is left is Bose statistics 
for identical  scalars, demanding total symmetry over their interchange, 
while the color wave function is antisymmetric. So
the only solution~\cite{PENTA:JafWil} is to make the spatial  wave
function antisymmetric by putting one of the diquark into the P-wave
state. It means that such pentaquarks should be 
degenerate with the excited P-wave decuplet baryons.
$ 
M_\Theta= 2\Sigma+\Sigma_s+\delta M_{L=1}+V_{residual}
$
where the first 2 terms are masses of the diquarks and strange quark,
plus an extra  contribution for the P-wave, plus whatever {\it residual}
interaction there might be.
One finds that the difference between P-wave and
S-wave state is $\delta M_{L=1}=\hbar \omega_\lambda\approx 480 \, MeV$
and thus
$ 
m_\Theta \approx m_\Sigma^*(3/2)+\delta M_{L=1}\approx 1400+480=1880 \,
{\rm MeV} \,\,,
$
which is  well above the  ``observed'' mass of 1540 MeV.

However, using one scalar and one {\it tensor} diquark one can do
without  the P-wave penalty, and the schematic mass estimate now reads
$  
m_\Theta \approx m_\Sigma^*(3/2)+\delta M_{T} \approx
 1400+150=1550\, {\rm MeV}\,\,, 
$
which is much closer to the experimental value.


Since the tensor diquark has the opposite parity, both possibilities
correspond to the same global parity $P=+1$. Also common to both
schemes is the fact that the total spin of 4 quarks is 1, so
adding the spin of the $\bar s$ can lead not only to $s= 1/2^+$ but also to
$s=3/2^+$ states (which are not yet observed). 

So, we conclude that if we only look at the {\it masses}, it appears that
it is better to substitute one diquark by its tensor variant, rather
than to enforce the P-wave. 
 Such an alternative scheme provides a different set of
 flavor representations,
 $\bar 3\otimes 6 \otimes \bar 3=1\oplus 8\oplus 8\oplus 10\oplus 27$.
The largest representation 27 has particles with
quantum numbers of $\Theta^+$ and $\Xi(1860)$, and even more exotic
triplets such as $\Omega$-like $sssq\bar q$ states.
The cascades have isospin 3/2, as observed. However $\Theta^+$
is a part of an isotriplet,
with $\Theta^{++}$ and $\Theta^0$ partners. The former can decay into
$pK^+$, a quite visible mode.

If one goes a step further, to 6-quark states,
for example for 3 $ud$ diquarks,
 the asymmetric color wave function asks for another
asymmetry: to do so one can put all 3 diquarks into the P-wave state,
with the spatial wave function $\epsilon_{ijk}\partial_i \subs \partial_j \subs \partial_k
\subs$  suggested in the second paper of \cite{DIQ:BoseCond}.
This will cost $3(\Sigma+\delta M_{L=1})=2700\, MeV$, 
well in agreement with the magnitude of the
repulsive nucleon-nucleon core.
However if one considers the quantum
numbers of the famous $H$ dibaryon, one can also make those out of diquarks
such as $\subs \subd \subu$. The resulting wave function is overall
flavor antisymmetric with all diquarks in S-states. Thus there is no
need for P-wave or tensor diquarks for the $H$ dibaryon. Our schematic model would then
lead to a light $H$ never seen.
This shows that
 additive schematic models ignore inter-diquark interaction:
the issue we will discuss in the next section.

\section{Diquark interaction via the correlation functions}
 In \cite{PS} we attempted to attack the multiquark dynamics directly,
propagating 4-6 quarks in 
the instanton liquid model. (Recall that the number of
quarks involved in the model space included in such calculations
is equal to the number of instanton zero modes,
equal to the number of instantons plus antiinstantons, which is
typically 200-300.)
   
We start with the operators for the
two diquark systems\footnote{Of course these systems carry
color and should be complemented by a heavy antiquark or another
diquark, which is factored out.} of all possible types, with
a variable splitting between two diquark operators at each time.
 For  spatially antisymmetric states the probability to
find both diquarks at the same point must vanish by Bose symmetry\footnote{In most lattice works, for example, the operators used so far
are only local, which prevents from approaching all $P$-wave states.}.
For each of the four possible cases we look at the dependence 
of the correlators on the number of flavors $\nf$, which translates
into the number of exchange diagrams. At $\nf = 4$ all four quarks
are different and there are no exchange diagrams, at $\nf = 3$ there is one, 
and at $\nf = 2$ we have two of them. Accordingly we would expect their
contribution to grow for decreasing $\nf$.


There are two ways to use the correlators: The usual one is go to the largest
$\tau$ (Euclidean time) possible and fit a mass from the logarithmic slope
 (very difficult). We use another one also, namely
fit the effective interaction at small $\tau$ 
for 2,3 diquarks placed at a variable distance.
 Some examples are shown in Fig.1(right).
 One can
clearly see a \emph{repulsive core}
at small distances $d$ of the diquarks, which gets stronger at smaller
number of quark flavors.
 The repulsion reaches roughly $300~\MeV$ and the width
of the core corresponds to approximately the
instanton radius $\rho = 0.35~\fm$. 

 \begin{figure}
 \includegraphics[width=7.cm]{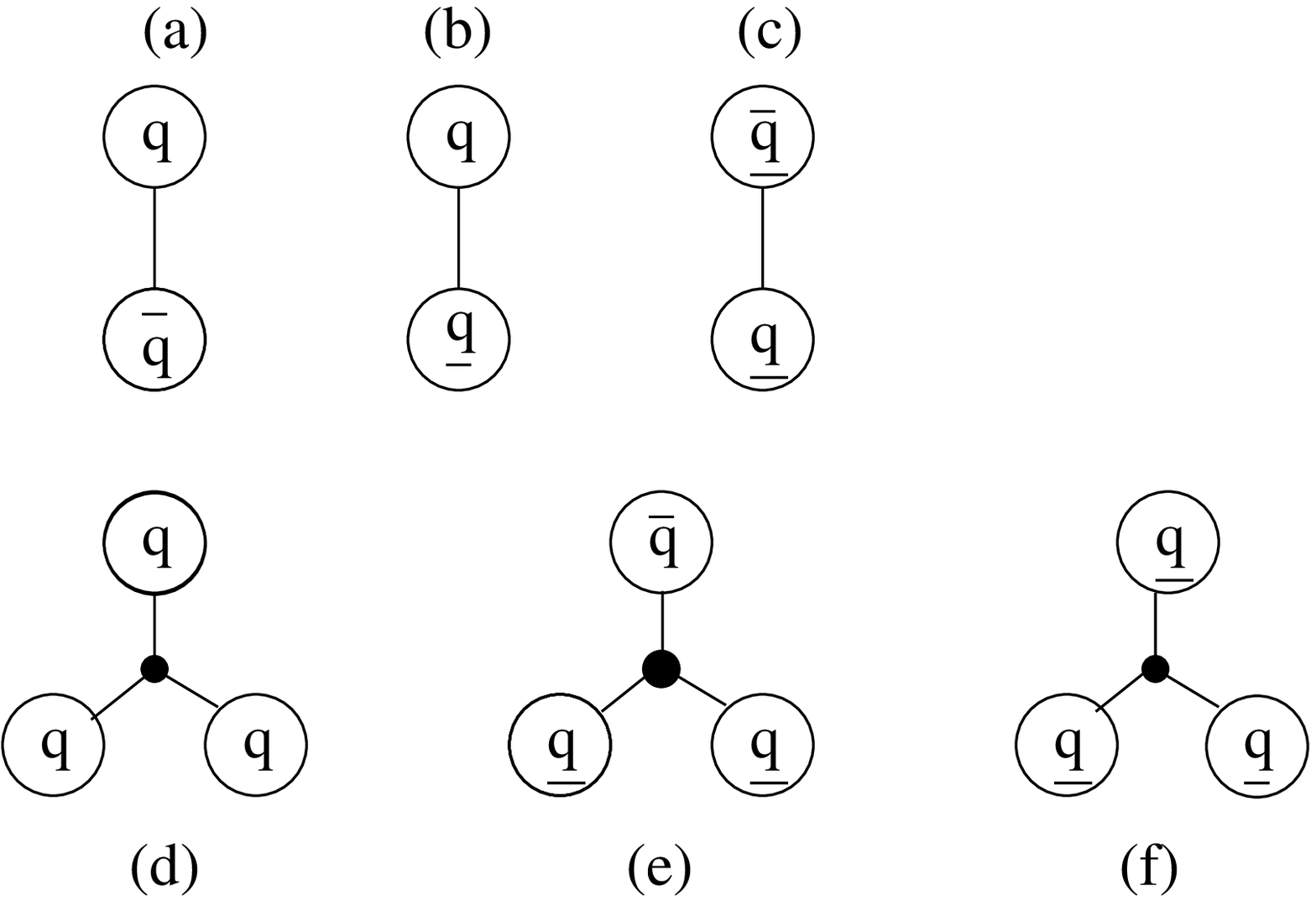}%
 \includegraphics[width=7.0cm]{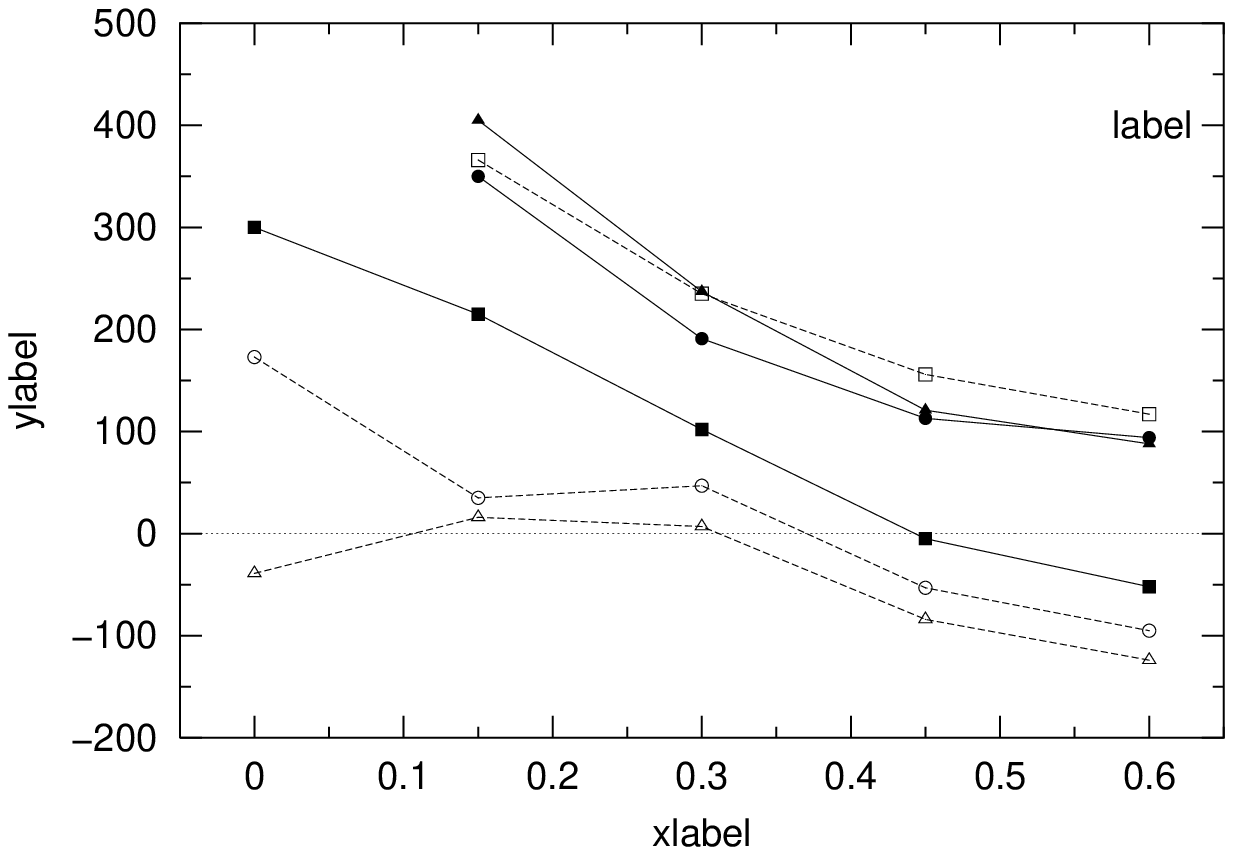}
 \caption{\label{plot:V.tetra}
(left) Schematic structure of (a)  ordinary mesons, (b) quark-diquark or
octet baryons, (c) diquark-antidiquark states or tetraquarks, (d) decuplet baryons, (e)
pentaquarks and (f) dibaryons.\\ 
(right)
 Effective interaction potential $V(d)$
 for the two diquark systems
 $\triC \triF \Sym$ $\nf=3,4$ ($ black square$);
 $\triC \sixF \Anti$ $\nf=2,3$ ($\bullet$), $\nf=4$ ($black triangle$);
 $\sixC \triF \Anti$ $\nf=3,4$ ($square$);
 $\sixC \sixF \Sym$ $\nf=2,3$ ($circle$), $\nf=4$ ($triangle$).
 The uncertainty in $V(d)$ is about $50~\MeV$.
}
\end{figure}
No quite light states other than meson+baryon were found so far for
pentaquarks,
and all operators with
with 3 scalar diquarks lead to
 flavor singlet H with  a large mass $\sim$ 3 GeV. Much more detailed
 studies
are to follow as the work is in progress.

\subsection{Acknowledgments}
The talk is based on a number of works done with
my collaborators, whom I should thank first.
In particular, discussion of the schematic models for pentaquarks
is done with Ismail Zahed and the study of multiquark correlators
with (a graduate student) Daniel Pertot. This work was partially supported by the US DOE grant DE-FG-88ER40388.

\end{document}